\documentstyle[prl,aps,floats,epsf,psfig]{revtex}
\draft
\def\nim#1#2#3  {{\em Nucl. Instr. Meth.} {\bf#1}, #2 (#3). }
\def\np#1#2#3   {{ Nucl. Phys.} {\bf#1}, #2 (#3). }
\def\pcps#1#2#3 {{ Proc. Cam. Phil. Soc.} {\bf#1}, #2 (#3). }
\def\pl#1#2#3   {{ Phys. Lett.} {\bf#1}, #2 (#3). }
\def\plc#1#2#3   {{ Phys. Lett.} {\bf#1}, #2 (#3); }
\def\prep#1#2#3 {{ Phys. Rep.} {\bf#1}, #2 (#3). }
\def\prev#1#2#3 {{ Phys. Rev.} {\bf#1}, #2 (#3). }
\def\prl#1#2#3  {{ Phys. Rev. Lett.} {\bf#1}, #2 (#3). }
\def\prs#1#2#3  {{ Proc. Roy. Soc.} {\bf#1}, #2 (#3). }
\def\ptp#1#2#3  {{ Prog. Th. Phys.} {\bf#1}, #2 (#3). }
\def\rmp#1#2#3  {{ Rev. Mod. Phys.} {\bf#1}, #2 (#3). }
\def\rpp#1#2#3  {{ Rep. Prog. Phys.} {\bf#1}, #2 (#3). }
\def\zp#1#2#3   {{ Z. Phys.} {\bf#1}, #2 (#3). }
\def\epj#1#2#3   {{ Eur. Phys. Jour.} {\bf#1}, #2 (#3). }

\begin{document}

\wideabs{
\title{Search for a 33.9 MeV/$c^2$ Neutral Particle in Pion Decay}
\author{
 J.~A.~Formaggio$^2$,
 E.~D.~Zimmerman$^2$,
 T.~Adams$^{4}$,
 A.~Alton$^{4}$,
 S.~Avvakumov$^7$,
 L.~de~Barbaro$^5$,
 P.~de~Barbaro$^7$
 R.~H.~Bernstein$^3$,
 A.~Bodek$^7$,
 T.~Bolton$^4$,
 J.~Brau$^6$,
 D.~Buchholz$^5$,
 H.~Budd$^7$,
 L.~Bugel$^{3}$,
 S.~Case$^2$,
 J.~M.~Conrad$^2$,
 R.~B.~Drucker$^6$,
 B.~T.~Fleming$^2$,
 R.~Frey$^6$,
 J.~Goldman$^4$,
 M.~Goncharov$^4$,
 D.~A.~Harris$^7$,
 R.~A.~Johnson$^1$,
 J.~H.~Kim$^2$,
 S.~Koutsoliotas$^2$,
 M.~J.~Lamm$^3$,
 W.~Marsh$^3$,
 D.~Mason$^6$,
 K.~S.~McFarland$^{7,3}$,
 C.~McNulty$^2$,
 D.~Naples$^4$,
 P.~Nienaber$^{3}$,
 A.~Romosan$^2$,
 W.~K.~Sakumoto$^7$,
 H.~M.~Schellman$^5$,
 M.~H.~Shaevitz$^2$,
 P.~Spentzouris$^{2,3}$,
 E.~G.~Stern$^2$,
 M.~Vakili$^1$,
 A.~Vaitaitis$^2$,
 V.~Wu$^{1}$,
 U.~K.~Yang$^7$,
 J.~Yu$^3$, and
 G~.P.~Zeller$^5$ 
}
\address{
$^1$ University of Cincinnati, Cincinnati, OH 45221 \\
$^2$ Columbia University, New York, NY 10027 \\
$^3$ Fermilab, Batavia, IL 60510 \\
$^4$ Kansas State University, Manhattan, KS 66506 \\
$^5$ Northwestern University, Evanston, IL 97403 \\
$^6$ University of Oregon, Eugene, OR 97403 \\
$^7$ University of Rochester, Rochester, NY 14627 \\
}
\date{\today}
\maketitle
\begin{abstract}
The E815 (NuTeV) neutrino experiment has performed a search for a
$33.9$~MeV/$c^2$ weakly-interacting neutral particle produced in
pion decay. Such a particle may be responsible for an anomaly in the 
timing distribution of neutrino interactions in the KARMEN experiment. 
E815 has searched for this particle's decays in an instrumented decay 
region; no evidence for this particle was found. The search is sensitive 
to pion branching ratios as low as $10^{-13}$.
\end{abstract}
\twocolumn
}

The KARMEN collaboration at the ISIS spallation neutron
facility at the Rutherford Appleton Laboratory uses a pulsed neutrino
beam resulting from stopped pion and muon decays to study
neutrino-nucleon interactions.  Their experiment has reported an
anomaly in the timing distribution of neutrino interactions from
stopped muon decays \cite{KARMEN}.  One possible explanation for the 
anomaly is an
exotic pion decay, where a neutral weakly-interacting or sterile particle is
produced and travels 17.7~m to the KARMEN detector with a velocity of
4.9 m/$\mu$s. Upon reaching the KARMEN detector, the exotic particle
decays to a partially electromagnetic state, such as $e^+e^-\nu$ or
$\gamma\nu$.  The $e^+ e^- \nu$ decay is strongly favored by recent
KARMEN data \cite{oehlers}. This slow moving exotic particle
(hereafter denoted as $Q^0$) would have a mass of 33.9 MeV/$c^2$,
which is near the kinematic threshold for pion decay.

The KARMEN experiment reports a signal curve for pion branching ratio
$B(\pi \rightarrow \mu+Q^0) \cdot B(Q^0 \rightarrow {\rm visible})$
versus lifetime.  Their signal region extends as low as $10^{-16}$ for
a lifetime of 3.6 $\mu$s.  For branching ratios above this minimum,
there exist two solutions to the KARMEN anomaly (at small and large
lifetimes). Certain portions of the KARMEN signal have already been
excluded. Experiments at PSI \cite{PSI,PSIEE} have
performed searches for this exotic particle by studying the momentum
spectrum of muons and electrons produced by $\pi^+$ decays in
flight. PSI has excluded any exotic pion decays to muons with
branching ratios above $2.1 \times 10^{-8}$ at 90\% C.L., and to
electrons with branching ratios above $0.9 \times 10^{-6}$ at 90\%
C.L. In addition, there exist astrophysical constraints on certain
decay modes of the $Q^0$ which exclude lifetimes above $10^3$~s
\cite{astro}.  Despite the above limits, portions of the KARMEN
allowed signal region remain to be addressed.
\begin{figure*}
\label{detector}
\centerline{
\psfig{figure=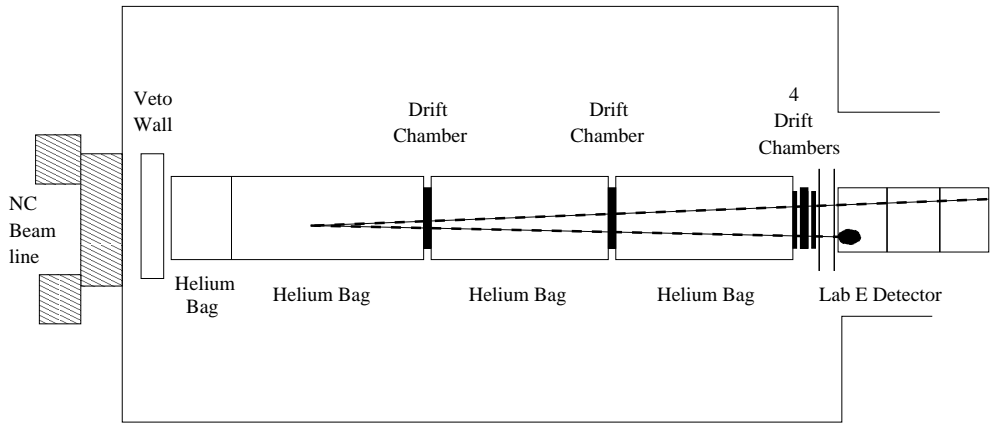,height=2.2in}}
\caption{Schematic of the NuTeV decay channel, including the veto
wall, helium bags, drift chambers, and calorimeter.}
\end{figure*}

The E815 (NuTeV) neutrino experiment at Fermilab has performed a
direct search for the $Q^0$ decay by combining the capabilities of a
high intensity neutrino beam with an instrumented decay region (the
``decay channel'').  During the 1996-1997 fixed target run at
Fermilab, NuTeV received 2.54$\times 10^{18}$ 800 GeV protons striking
a BeO target with the detector configured for this search.  The
secondary pions and kaons produced from the interaction were
subsequently sign-selected using a series of magnets and focused down
a beamline at a 7.8 mrad angle from the primary proton beam direction.
The pions and kaons could then decay in a 440 m pipe before hitting a
beam dump. A total of $(1.4 \pm 0.1) \times 10^{15}$ pion decays and
$(3.6 \pm 0.4) \times 10^{14}$ kaon decays occurred in the pipe. The
neutral weakly-interacting decay products (neutrinos and possibly
$Q^0$'s) traveled through approximately 900 meters of earth berm
shielding before arriving at the decay channel.

The instrumented decay channel consisted of a series of helium bags,
extending a total of 34 meters in length, interspersed with 3 m
$\times$ 3 m multi-wire argon-ethane drift chambers. The drift
chambers were designed to track charged particles from decays
occurring within the helium.  Upstream of the decay channel stood a
4.6 m $\times$ 4.6 m array of scintillation plates, known as the veto
wall, used to detect any charged particles entering from upstream of
the detector.  Downstream of the decay channel was the Lab E neutrino
detector, which consisted of a 690-ton iron-scintillator sampling
calorimeter interspersed with drift chambers. The Lab E detector
provided triggering, energy measurement, and final particle
identification for tracks entering from the decay channel. Particles
were identified by their penetration into the calorimeter: a muon
produced a long track, a pion produced an elongated cluster of hits,
and an electron or photon produced a compact cluster. More details of
the decay channel and the calorimeter can be found elsewhere
\cite{NHL,NuTeV}.

The decay channel was employed to search for the decay of the $Q^0$. 
The experimental signature of this decay is a low-mass, low-transverse 
momentum electron-positron pair having a vertex within the fiducial 
volume of the decay channel.

A series of analysis cuts isolated the $Q^0$ from background.  The
cuts were divided into two categories: reconstruction and kinematic.
Reconstruction cuts isolated two track events occurring within the
fiducial volume of the decay channel.  We required that two charged
tracks originated from a common vertex, with no additional tracks
associated with the vertex.  By removing events with activity in the
veto wall, we ensured that no charged tracks entered from upstream of
the decay channel. We also required that each track had a small slope
($< 10$ mr) relative to the beam axis; and that, when projected
upstream to the veto wall, it fell within 50'' of the beam
center. These cuts removed both cosmic rays and photons from neutrino
interactions in the upstream berm.  Finally, we required that each
track be identified as an electron based on the shower shape in the
calorimeter.  Because of the small opening angle of the two tracks,
the two electron showers manifested themselves as a single merged
electron-like cluster within the calorimeter.  The efficiency for
identifying such an event as an $e e$ pair was estimated from Monte
Carlo studies to be 90.1 \%.  The cluster energy was divided between
the tracks based on a fit to the amount of multiple scattering each
track underwent in the decay channel.

\begin{figure}
\centerline{
\psfig{figure=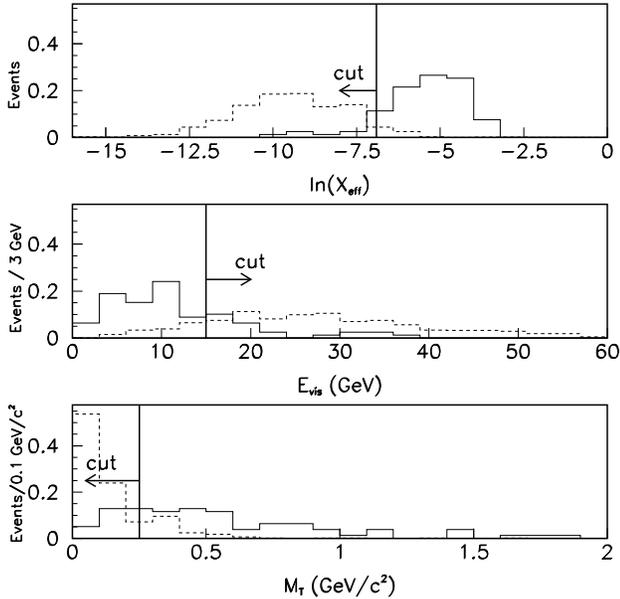,height=3.5in}}
\caption{Kinematic Monte Carlo distributions of $x_{\rm eff}$, visible
energy, and transverse mass for neutrino interactions (solid) and
$Q^0$ (dashed).  All events shown have passed reconstruction cuts only
and are relatively normalized.
\label{kine}}
\end{figure}

Because of its low mass, the $Q^0$ possesses unique kinematic features
which can be used to distinguish it from potential background sources,
such as photons and deep-inelastic neutrino interactions. We have used
effective scaling variables to represent the kinematics of the
reconstructed events. The effective scaling variables $x_{\rm eff}$
and $W_{\rm eff}$ were calculated for each event using the following
assumptions: 1) the event was a charged current neutrino interaction
($\nu_{\ell} N \rightarrow \ell N'X$) and 2) the missing transverse
momentum in the event was carried by an undetected final state
nucleon.  We have defined $x_{\rm eff} \equiv \frac{Q^2_{\rm
vis}}{2m_p\nu_{\rm vis}}$ and $W_{\rm eff} \equiv \sqrt{m_p^2 + 2m_p
\nu_{\rm vis}/c^2 - Q^2_{\rm vis}/c^2}$, where $Q_{\rm vis}$ is the
visible reconstructed 4-momentum transfer, $\nu_{\rm vis}$ is the
reconstructed hadron (or electron) energy, and $m_p$ is the proton
mass.  Using the above definitions, we required that all reconstructed
events have $x_{\rm eff} < 0.001$ and $W_{\rm eff} > 2.5$ GeV/c$^2$.
In addition, we required that the reconstructed transverse mass $m_T$
($m_T \equiv |p_T| + \sqrt{p_T^2+m_{\rm V}^2}$, where $m_{\rm V}$ is
the invariant mass of the two charged tracks and $p_T$ is the momentum
transverse to the beam axis) be less than 250 MeV/c$^2$. Finally, we
required that the total energy deposited by the $e^+e^-$ pair be
greater than 15 GeV. The effect of these cuts when applied to signal
and typical background kinematic distributions can be seen in
Figure~\ref{kine}. The total acceptance for $Q^0$ events in the
fiducial region was 15.6\%.

The principal backgrounds originated from three main sources: neutrino
interactions in the helium, neutrino interactions in the drift
chambers, and neutral particles (mainly photons and kaons) from
neutrino interactions in the berm and veto wall.  Note that because
the two electron tracks had a very small opening angle, the
longitudinal vertex position resolution was quite poor ($\sigma
\approx 7$ m).  Thus, interactions in the chambers could not be
removed using a vertex position cut. We used the Lund Monte Carlo
program \cite{LUND} to simulate the $\nu p$ deep-inelastic
interactions in both the berm and the decay channel. Separate Monte
Carlo programs were used to simulate hadronic resonance and
diffractive single pion production ~\cite{rein,diff}.  The GEANT Monte
Carlo program (version 3.21) \cite{GEANT} was used to simulate the
decay channel and Lab E calorimeter.  Based on our Monte Carlo study,
we expected $0.06 \pm 0.05$ total background events within our signal
region (See Table~\ref{background}).  The uncertainty on this
background estimate is dominated by Monte Carlo statistics.

\begin{table}
\caption{ Total Expected Background Events. Errors reflect Monte Carlo
statistics. \label{background}}
\begin{tabular}{|c|c|}
Source & Rate \\
\hline
\\
Photons & $0.04 ^{+0.02}_{-0.04} $ \\ 
Kaons & $\ll$ 0.001 \\
Deep Inelastic Charged Current & 0.00 $\pm$ 0.04 \\  
Deep Inelastic Neutral Current & 0.02 $\pm$ 0.02 \\  
Cosmic rays & $\ll$ 0.001 \\
Quasi-Elastic Charged Current & 0.000 $\pm$ 0.008 \\  
Resonance Neutral Current & 0.000 $\pm$ 0.003 \\  
Diffractive Pions & $\pm$ 0.01 \\
\hline
Total & 0.06 $\pm$ 0.05 \\
\end{tabular}
\end{table}

\begin{figure}
\centerline{
\psfig{figure=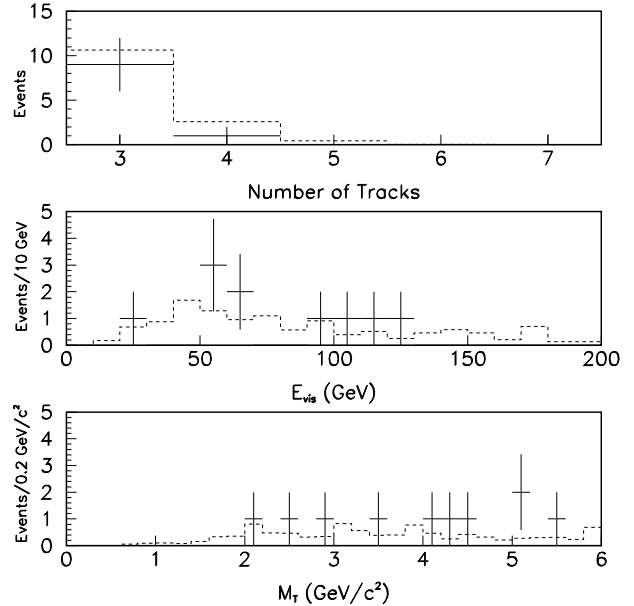,height=3.5in}}
\caption{Kinematic distributions of multiplicity, energy, and
transverse mass for data (crosses) and background Monte Carlo (dashed)
multi-track events.  Monte Carlo is absolutely
normalized. \label{antibox4}}
\end{figure}

A blind analysis was performed. The signal region was hidden while
cuts were developed based on Monte Carlo studies of signal
efficiencies and background rejection. Before examining the data near
or in the signal region, we performed a series of studies to verify
our background estimates. Using the Monte Carlo, we made predictions
for the following three quantities: 1) the number of low energy (below
15 GeV) and high transverse mass (above 500 MeV/c$^2$) background
events; 2) the number of $\mu\pi$ events; and 3) the number of
multi-track events occurring within our decay channel. The results of
these studies (shown in Table~\ref{antibox}) demonstrate good
agreement between data and the Monte Carlo predictions.  In the case
of multi-track events, where a larger sample of events was available,
there was also good agreement for various kinematic distributions (see
\begin{table}
\caption{Background Study Results. Uncertainties are
systematic.\label{antibox}}
\begin{tabular}{|c|c|c|}
Type of Event & Events Predicted & Events Seen \\
\hline
High Transverse Mass & 2.0 $\pm$ 0.3 & 1 \\
$\mu \pi$ Events & 4.1 $\pm$ 0.6 & 3 \\
Multiple Track Events & 13.7 $\pm$ 1.8 & 10 \\
\end{tabular}
\end{table}
Figure~\ref{antibox4}).

The above acceptance estimate was based on a heavy neutrino
model where the $Q^0$ decays to $e e \nu$ \cite{GLR}.  We also considered
additional decay models where the $Q^0$ decays to $\gamma \nu$ and
$\gamma \gamma \nu$. The acceptances under these decay scenarios were
$(0.5 \pm 0.1)\%$ and $(1.1\pm 0.1)\%$, respectively.  The low
efficiencies are due to the requirement that two tracks be
reconstructed; this was only possible if a photon converted in the
low-mass decay region before entering the calorimeter.

Systematic errors for this result were dominated by uncertainties on
the number of pion decays in the pipe ($6.8\%$) and the overall
normalization. The sensitivity normalization was taken from a
measurement of the number of neutrino interactions in the decay
channel using very loose cuts. This number was $(10.0 \pm 4.3$)\%
below a prediction normalized to the number of neutrino interactions
in the calorimeter. Considering this disagreement as a systematic
error on the normalization, we have calculated a total systematic
error of $12.1$\% on the sensitivity.

\begin{figure}
\centerline{
\psfig{figure=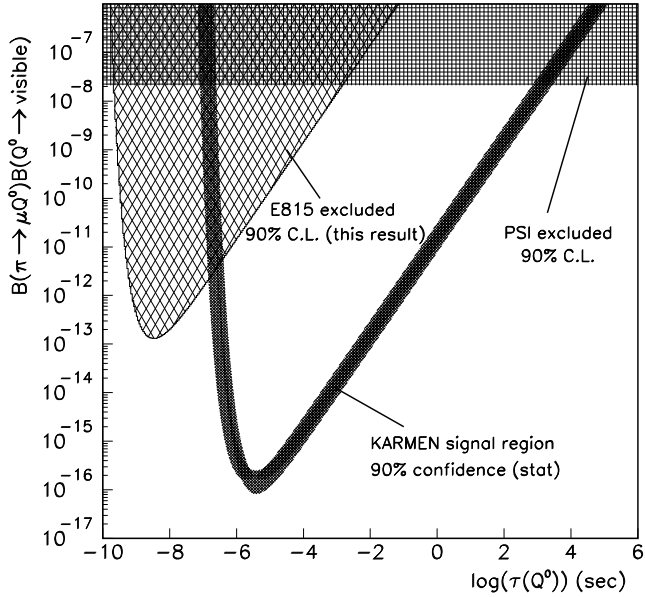,height=3.5in}}
\caption{Branching ratio versus lifetime plot for the Karmen signal
and the exclusion regions ( 90\% C.L.) from the NuTeV and PSI
experiments.  Systematic errors (except for decay model) have been
included. \label{Qmu}}
\end{figure}

\begin{figure}
\centerline{
\psfig{figure=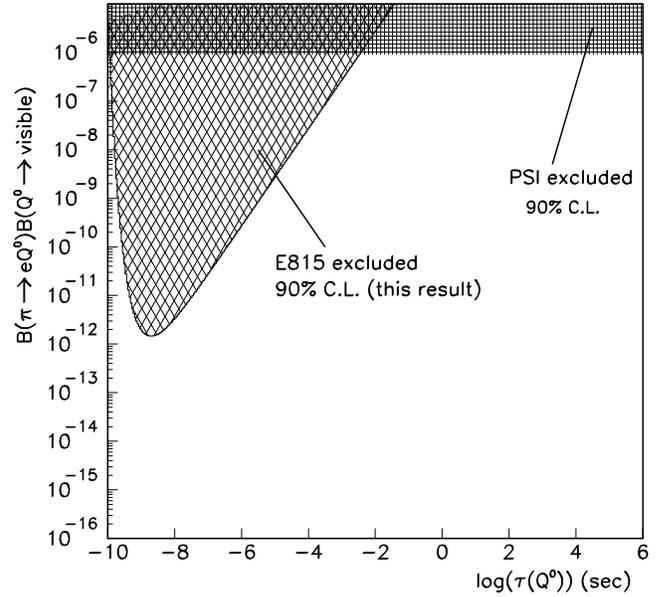,height=3.5in}}
\caption{Branching ratio versus lifetime plot for the NuTeV limit for
$\pi \rightarrow e + Q^0$.  Systematic errors (except for decay model)
have been included. \label{Qee}}
\end{figure}

Upon analyzing the signal region, we found no events which passed the
selection criteria.  The probability of seeing zero events from an
expected background of $0.06\pm 0.05$ is 94\%.  We thus present an
upper limit, shown in Fig.~\ref{Qmu}, on $B(\pi \rightarrow \mu+Q^0)
\cdot B(Q^0 \rightarrow {\rm visible})$.\footnote{ A simple
interpretation of the $Q^0$ is that the particle is a sterile heavy
neutrino which mixes with the muon neutrino.  This requires a
branching ratio $B(\pi\rightarrow \mu + Q^0) \approx 6\times 10^{-8}$,
which has already been ruled out by the PSI limit \cite{Govaerts}.
Still viable is a model where the heavy neutrino is produced by a
smaller mixing with $\nu_\mu$ and decays primarily via a larger mixing
with $\nu_\tau$, such that $|\langle Q^0| \nu_{\mu}
\rangle|\cdot|\langle Q^0| \nu_{\tau} \rangle| \approx 2 \times
10^{-6}$.  Within this model, we set a limit on this product $|\langle
Q^0| \nu_{\mu} \rangle|\cdot|\langle Q^0| \nu_{\tau} \rangle| < 1.4
\times 10^{-3}$ at 90\% C.L.  } NuTeV is also sensitive to $\pi
\rightarrow e + Q^0$; the limit is shown in Fig.~\ref{Qee}.


This result excludes a region of parameter space which extends as low 
as four orders of magnitude below current limits on the short lifetime
solution to the KARMEN anomaly. An experiment with significantly more
pion decays will be necessary to confirm or rule out the longer lifetime 
and lower branching ratio regions.

This research was supported by the U.S. Department of Energy and the
National Science Foundation.  We thank the staff of FNAL for their
contributions to the construction and support of this experiment
during the 1996-1997 fixed target run.



\begin{references}
\vspace{-.30in} 
%
\bibitem{KARMEN} B. Zeitnitz, in Proc. 17th International Workshop on
Weak Interactions and Neutrinos (1999).
%
\bibitem{oehlers} C. Oehler, in Proc. Sixth Topical Seminar on
Neutrino and Astroparticle Physics (1999).
%
\bibitem{PSI} 
M. Daum {\it et al.}, \pl{B361}{179}{1995}
%
\bibitem{PSIEE} 
N. De Leener-Rosier {\em et al.}, \pl{D43}{3611}{1991}
%
\bibitem{astro}
D. Choudhury {\em et al.}, e-print hep-ph/9911365.
%
\bibitem{NHL} 
A. Vaitaitis {\em et al.}, \prl{83}{4943}{1999}
%
\bibitem{NuTeV} 
D. A. Harris {\em et al.}, e-print hep-ex/9908056.
%
\bibitem{LUND}
G. Ingelman {\em et al.}, {\em Comput. Phys. Commun.} {\bf 101}, 108 (1997).
%
\bibitem{rein}
D.Rein and L.M. Sehgal, {\em Ann. Phys.} {\bf 133}, 71 (1981).
%
\bibitem{diff}
T. Adams {\em et al.}, e-print hep-ex/9909041.
%
\bibitem{GEANT}
CERN CN/ASD, GEANT detector description and simulation library (1998).
%
\bibitem{GLR}
M.Gronau, C.N. Leung, and J.L. Rosner, \prev{D29}{2539}{1984}
%
\bibitem{Govaerts}
J. Govaerts, J. Deutsch, and P.M. Van Hove, \pl{B389}{700}{1996}
%
\end{references}
\end{document}